\documentclass[10pt,conference]{IEEEtran}
\usepackage[T1]{fontenc}
\usepackage[utf8]{inputenc}
\usepackage{cite}
\usepackage{amsmath,amssymb}
\usepackage{booktabs}
\usepackage{listings}
\usepackage{balance}
\usepackage[hidelinks]{hyperref}
\usepackage{xcolor}
\definecolor{forestgreen}{RGB}{34,139,34}
\usepackage{booktabs}
\usepackage{tabularx}
\usepackage{array}
\lstdefinelanguage{Rust}{
  keywords={fn, let, mut, return, if, else, while, for, in, loop,
  match, struct, enum, impl, trait, pub, use, requires, ensures, assert},
  comment=[l]{//},
  morecomment=[s]{/*}{*/},
}
\newcommand{\benchmarkName}{VerusWeave}
\hypersetup{
  pdftitle={\benchmarkName: A Benchmark for Proof Repair in Concurrent Verus Programs},
  pdfauthor={Pedro Ribeiro; Martim Afonso; André Vasconcelos; João F. Ferreira},
  pdfkeywords={benchmark, proof repair, concurrent software, Verus, coding agents}}

\newcommand{\dataseturl}{\texttt{https://doi.org/10.5281/zenodo.23540079}}
\newcommand{\videourl}{\texttt{https://youtu.be/cOC\_ilmSATY}}

\begin{document}

\title{\benchmarkName: A Benchmark for Proof Repair in Concurrent Verus Programs}

\author{\IEEEauthorblockN{Pedro Ribeiro\IEEEauthorrefmark{1},
  Martim Afonso\IEEEauthorrefmark{1},
  Andr\'e Vasconcelos\IEEEauthorrefmark{1},
  Jo\~ao F. Ferreira\IEEEauthorrefmark{2}}
\IEEEauthorblockA{\IEEEauthorrefmark{1}INESC-ID and Instituto Superior T\'ecnico, Universidade de Lisboa, Portugal}
\IEEEauthorblockA{\IEEEauthorrefmark{2}INESC-ID and Faculty of Engineering, University of Porto, Portugal}
%\IEEEauthorblockA{\texttt{\{pedro.melo.ribeiro,martim.afonso,andre.vasconcelos\}@tecnico.ulisboa.pt, joao@joaoff.com}}
%\pedroemail, \martimemail, \andreemail, \joaoemail}}
}
\maketitle

\begin{abstract}

Concurrent software underpins critical infrastructure, including operating systems and databases, yet thread interactions can have bugs that bypass conventional testing. Verus extends Rust with formal verification capabilities, enabling developers to specify, implement and prove the correctness of concurrent code in the same language. Although coding agents have been explored for automated proof construction, existing Verus benchmarks offer limited coverage of concurrency-specific examples that connect atomic operations to abstract protocols.

To address this gap, we introduce \benchmarkName{}, the first benchmark dedicated to concurrent proof repair, derived from four open-source projects. It comprises 75 focused tasks and 17 full-module variants that exercise features used to verify low-level concurrent code.
Tasks remove selected proof annotations and require agents to fill gaps, while preserving executable code and specifications. The evaluation harness validates repairs using Verus and a change checker, recording solutions, execution traces, and resource usage. An initial evaluation of four general-purpose coding-agent harnesses achieved 75--85\% success on focused tasks and 29--47\% on full-module variants, highlighting the challenge of autonomous module-wide repairs. \benchmarkName{} provides a reusable foundation for evaluating automated proof construction for concurrent software.

Dataset: \dataseturl.

Video: \videourl.
\end{abstract}


\section{Introduction}

Large Language Models (LLMs) and autonomous programming agents are transforming software development, but generated code is not necessarily correct, and its growing volume increases verification demands. Providing stronger correctness guarantees without prohibitive development effort is therefore increasingly important.

Formal verification addresses this need by mathematically establishing that programs satisfy their specifications, unlike testing, which examines finitely many inputs. However, its adoption remains limited by the expertise, cost, and maintenance effort required to construct and evolve proofs~\cite{brugger2025practicaldeductiveverificationinsights,oliveira2025challenges}. LLMs can automate specification generation, invariant inference, and proof construction~\cite{leveraging_llms_dafny,silva2025inferring,rango,autorocq}, while integrating verification into coding agents could further ease the development and maintenance of verified software.

Concurrent software presents a particularly challenging target for this automation. It underpins critical infrastructure, including operating-system kernels, databases, and distributed services, yet bugs may depend on rare thread interleavings that testing struggles to expose~\cite{chess}. Proving correctness also requires reasoning about inter-thread interference, shared-state invariants, and memory-access permissions, making concurrency a demanding setting for automated proof generation.

Verus~\cite{lattuada2023verusverifyingrustprograms} provides a foundation for addressing this challenge by extending Rust with specifications and proof constructs for machine-checked verification. Its applications to operating systems and security modules demonstrate the practical relevance of this approach~\cite{cortenmm,verismo,owlc}. Meanwhile, coding agents can iteratively modify code, invoke tools, and respond to compiler and verifier feedback. This makes proof repair a natural target for automation: agents can refine proof annotations until verification succeeds, without changing the executable implementation or specification. Successful verification provides machine-checkable evidence of correctness relative to the encoded specification and verifier assumptions.

Despite growing interest in LLM-assisted verification, existing Verus benchmarks provide limited coverage of the low-level concurrency primitives and proof patterns needed to verify concurrent systems. Consequently, coding agents' ability to repair such proofs, the difficulties they encounter, and the effects of agent and harness design remain underexplored.

% To address this gap, we introduce \benchmarkName{}, the first benchmark for evaluating agents on concurrent Verus proof repair. Derived from existing Verus projects, it comprises 75 focused tasks and 17 full-module tasks exercising low-level concurrency primitives. VerusWeave targets researchers developing automated proof-repair methods and studying their effectiveness, failure modes, and resource usage. Tasks are distributed as standalone Rust files and we provide an evaluation harness that records verification outcomes, resource statistics, and solutions through run-level JSON files. Logs, statistics and solutions from an initial experimental evaluation ship alongside the dataset. 

To address this gap, we introduce \benchmarkName{}, a benchmark for evaluating agents on concurrent Verus proof repair. Derived from existing Verus projects, it comprises 75 focused tasks and 17 full-module variants exercising low-level concurrency primitives. VerusWeave targets researchers developing automated software repair methods and studying agents effectiveness, failure modes, and resource usage. The dataset contains standalone Rust tasks alongside logs, statistics, and solutions from an initial evaluation. We include an evaluation harness to run the benchmark, validate candidates and inspect run-level JSON records containing validation outcomes, resource statistics, and solution paths.

\section{Benchmark Design and Construction}

The benchmark targets two proof-obligation families in Verus's low-level concurrency support~\cite{hance2024verifying}: ghost-resource manipulation around atomic operations and inductiveness proofs for tokenized state machines. These families address different, but complementary, parts of a concurrent correctness argument. The first connects an executable synchronization event to a change in abstract state; the second establishes that the abstract change is permitted by the protocol.

% The former connects executable atomic operations to abstract protocol transitions; the latter proves that transitions preserve state-machine invariants.

\subsection{Task Families}

% \begin{figure*}[!t]
%   \centering
%   \begin{minipage}[t]{0.48\textwidth}
%     \vspace{0pt}
%     \lstinputlisting[
%       language=Rust,
%       basicstyle=\fontsize{5}{5.5}\selectfont\ttfamily,
%       aboveskip=0pt,
%       belowskip=0pt,
%       breaklines=true,
%       columns=fullflexible,
%       escapeinside={(*@}{@*)}
%     ]{tables_and_code/benchmark_once_get_highlighted_new.rs}
%     \caption{Atomic-operation task from \texttt{OnceImpl::get}.
% The atomic load and executable code remain unchanged.
% The added ghost
% recovers the permission to access the cell when its state
% indicates that initialization is complete. This permission
% allows the subsequent code to safely borrow the stored value.}
%     \label{fig:atomicTaskExample}
%   \end{minipage}
%   \hfill
%   \begin{minipage}[t]{0.48\textwidth}
%     \vspace{0pt}
%     \lstinputlisting[
%       language=Rust,
%       basicstyle=\fontsize{5}{5.5}\selectfont\ttfamily,
%       aboveskip=0pt,
%       belowskip=0pt,
%       breaklines=true,
%       columns=fullflexible,
%       escapeinside={(*@}{@*)}
%     ]{tables_and_code/benchmark_rwlock_inductive_highlighted_new.rs}
%     \caption{State-machine task for a reader--writer lock.
% The missing proof shows
% that the lock starts in a consistent state: no readers have
% requested or obtained access, and all reader counts are zero.}
%     \label{fig:inductiveTaskExample}
%   \end{minipage}
% \end{figure*}

\begin{figure}[t]
  \centering
\begin{lstlisting}[
    language=Rust,
    basicstyle=\fontsize{5}{5.5}\selectfont\ttfamily,
    aboveskip=0pt,
    belowskip=0pt,
    breaklines=true,
    columns=fullflexible,
    escapeinside={(*@}{@*)}
  ]
pub fn get<'a>(&'a self) -> (r: Option<&'a V>)
    requires
        self.wf(),
    ensures
        self.wf(),
        r matches Some(res) ==> self.inv().inv(*res),
{
    let tracked mut points_to = None;
    let res = atomic_with_ghost! {
        &self.state => load(); ghost g => {
(*@\textcolor{forestgreen}{\ttfamily\detokenize{+ }\hspace*{10\fontcharwd\font`M}\detokenize{match g }\{}@*)
(*@\textcolor{forestgreen}{\ttfamily\detokenize{+ }\hspace*{14\fontcharwd\font`M}\detokenize{OnceState::Init(points_to_opt) => }\{}@*)
(*@\textcolor{forestgreen}{\ttfamily\detokenize{+ }\hspace*{18\fontcharwd\font`M}\detokenize{points_to = Some(points_to_opt);}}@*)
(*@\textcolor{forestgreen}{\ttfamily\detokenize{+ }\hspace*{14\fontcharwd\font`M}\}}@*)
(*@\textcolor{forestgreen}{\ttfamily\detokenize{+ }\hspace*{14\fontcharwd\font`M}\detokenize{_ => }\{\}}@*)
(*@\textcolor{forestgreen}{\ttfamily\detokenize{+ }\hspace*{10\fontcharwd\font`M}\}}@*)
        }
    };
    // ... remaining code omitted
}
\end{lstlisting}
  \caption{Atomic-operation task from \texttt{OnceImpl::get}.
  The added ghost code recovers the permission to
  the initialized cell; executable code is unchanged.}
  \label{fig:atomicTaskExample}
\end{figure}

\begin{figure}[t]
  \centering
\begin{lstlisting}[
    language=Rust,
    basicstyle=\fontsize{5}{5.5}\selectfont\ttfamily,
    aboveskip=0pt,
    belowskip=0pt,
    breaklines=true,
    columns=fullflexible,
    escapeinside={(*@}{@*)}
  ]
#[inductive(initialize)]
fn initialize_inductive(post: Self, rc_width: int, init_t: T) {
(*@\textcolor{forestgreen}{\ttfamily\detokenize{+ }\hspace*{2\fontcharwd\font`M}\detokenize{assert forall |r| 0 <= r < post.rc_width implies}}@*)
(*@\textcolor{forestgreen}{\ttfamily\detokenize{+ }\hspace*{6\fontcharwd\font`M}\#\detokenize{[trigger] post.ref_counts.index(r) ==}}@*)
(*@\textcolor{forestgreen}{\ttfamily\detokenize{+ }\hspace*{10\fontcharwd\font`M}\detokenize{post.shared_pending.count(r) as int +}}@*)
(*@\textcolor{forestgreen}{\ttfamily\detokenize{+ }\hspace*{10\fontcharwd\font`M}\detokenize{Self::filter_r(post.shared_guard, r).len() as int}}@*)
(*@\textcolor{forestgreen}{\ttfamily\detokenize{+ }\hspace*{2\fontcharwd\font`M}\detokenize{by }\{}@*)
(*@\textcolor{forestgreen}{\ttfamily\detokenize{+ }\hspace*{6\fontcharwd\font`M}\detokenize{assert(post.ref_counts.index(r) == 0);}}@*)
(*@\textcolor{forestgreen}{\ttfamily\detokenize{+ }\hspace*{6\fontcharwd\font`M}\detokenize{assert(post.shared_pending.count(r) == 0);}}@*)
(*@\textcolor{forestgreen}{\ttfamily\detokenize{+ }\hspace*{6\fontcharwd\font`M}\detokenize{assert_multisets_equal!(}}@*)
(*@\textcolor{forestgreen}{\ttfamily\detokenize{+ }\hspace*{10\fontcharwd\font`M}\detokenize{Self::filter_r(post.shared_guard, r),}}@*)
(*@\textcolor{forestgreen}{\ttfamily\detokenize{+ }\hspace*{10\fontcharwd\font`M}\detokenize{Multiset::empty(),}}@*)
(*@\textcolor{forestgreen}{\ttfamily\detokenize{+ }\hspace*{6\fontcharwd\font`M}\detokenize{);}}@*)
(*@\textcolor{forestgreen}{\ttfamily\detokenize{+ }\hspace*{6\fontcharwd\font`M}\detokenize{assert(Self::filter_r(post.shared_guard, r).len() == 0);}}@*)
(*@\textcolor{forestgreen}{\ttfamily\detokenize{+ }\hspace*{2\fontcharwd\font`M}\}}@*)
(*@\textcolor{forestgreen}{\ttfamily\detokenize{+ }\hspace*{2\fontcharwd\font`M}\detokenize{assert(post.shared_counts_agree());}}@*)
}
\end{lstlisting}
  \caption{Reader--writer lock state-machine task.
  The added proof establishes zero reader counts
  and no pending or active readers at initialization.}
  \label{fig:inductiveTaskExample}
\end{figure}

% \emph{\textbf{Atomic-operation tasks}} remove ghost code associated with a fixed atomic operation, such as load, compare-and-exchange, or fetch-and-add. Ghost code tracks abstract protocol state and resource ownership without affecting execution, while invariants relate this information to the shared atomic value. The removed code lies within an \texttt{atomic\_with\_ghost!} block or another proof context permitting ghost-resource updates.
% The agent must reason about the atomic value before and after the operation and its return value to determine the necessary updates. A task is complete when the ghost updates are shown to preserve the invariants.

\emph{\textbf{Atomic-operation tasks}} require the agent to restore ghost code associated with a fixed atomic operation, such as load, compare-and-exchange, or fetch-and-add. This code tracks abstract protocol state and resource ownership without affecting execution; invariants connect that state and ownership to the shared atomic value. The omitted code appears within an \texttt{atomic\_with\_ghost!} block or another proof context that permits ghost-resource updates. 
To reconstruct it, the agent must reason about the operation’s effect on the atomic value and its return value to determine how to update the ghost state.
A task is complete when the restored ghost updates are verified to preserve the invariants.

\emph{\textbf{State-machine tasks}} require the agent to restore an inductiveness proof for an abstract concurrent protocol defined with Verus's \texttt{tokenized\_state\_machine!} macro, that specifies the program's possible states, initialization, transitions, invariants, and ghost-resource ownership. Each task retains these definitions but removes the inductive
% \texttt{\#[inductive(...)]} 
proof lemma for a selected transition. The agent must show that, if the invariants hold before the transition, they also hold afterward.

% \emph{\textbf{State-machine tasks}} target abstract concurrent protocols defined with Verus's \texttt{tokenized\_state\_machine!} macro, which specifies possible states, initialization, transitions, invariants, and ghost-resource ownership. Tasks retain these definitions but remove the \texttt{\#[inductive(...)]} proof lemma for a selected transition. The agent must prove that the transition preserves the state-machine's invariants, ensuring protocol consistency.

Figures~\ref{fig:atomicTaskExample} and~\ref{fig:inductiveTaskExample} illustrate representative tasks from the atomic-operation and state-machine families, respectively.

\subsection{Source Projects and Selection}

The corpus draws on open-source Verus code from Verified Memory Allocator~\cite{verified_memory_allocator}, Verified Node Replication~\cite{verified_node_replication}, and \texttt{vostd}~\cite{vostd_repo}, the verification library used by Asterinas~\cite{asterinas}. Due to the scarcity of public projects exercising the target concurrency mechanisms, we also include examples from the Verus repository~\cite{verus_repo}. These sources cover logical atomicity demonstrations, reader--writer locks, concurrent queues, memory allocation, and replicated logs. Table~\ref{tab:corpus} summarizes source provenance and task counts.

The \texttt{wait} task proof was developed during this work and
subsequently contributed to \texttt{vostd}, where it was merged in pull
request~702.\footnote{\href{https://github.com/asterinas/vostd/pull/702}{Asterinas/vostd
pull request 702}.}

\begin{table}[!t]
\caption{Source modules and repair tasks in \benchmarkName{}.
A: atomic-operation tasks; T: state-machine inductiveness tasks;
F: full-module variant. Module names omit \texttt{.rs}.}
\label{tab:corpus}
\centering
\small
\setlength{\tabcolsep}{4pt}
\renewcommand{\arraystretch}{0.92}
\begin{tabular}{@{}lrrc@{}}
\toprule
\textbf{Module} & \textbf{A} & \textbf{T} & \textbf{F} \\
\midrule
\multicolumn{4}{@{}l}{\emph{Verified Node Replication}} \\
context          & 2 & 0 & $\checkmark$ \\
cyclicbuffer     & 0 & 6 & $\checkmark$ \\
flat\_combine    & 0 & 8 & $\checkmark$ \\
log              & 5 & 0 & $\checkmark$ \\
replica          & 4 & 0 & $\checkmark$ \\
unbounded\_log   & 0 & 6 & $\checkmark$ \\
\addlinespace[2pt]
\multicolumn{4}{@{}l}{\emph{Verus examples}} \\
count\_to\_two   & 2 & 0 & $\checkmark$ \\
cuckoo           & 3 & 1 & $\checkmark$ \\
dist\_rwlock     & 0 & 4 & $\checkmark$ \\
fifo             & 2 & 5 & $\checkmark$ \\
helping          & 3 & 0 & $\checkmark$ \\
statics          & 1 & 0 & -- \\
\addlinespace[2pt]
\multicolumn{4}{@{}l}{\emph{Verified Memory Allocator}} \\
free             & 2 & 0 & $\checkmark$ \\
linked\_list     & 6 & 0 & $\checkmark$ \\
tokens           & 0 & 3 & $\checkmark$ \\
types            & 2 & 0 & $\checkmark$ \\
\addlinespace[2pt]
\multicolumn{4}{@{}l}{\emph{VOSTD}} \\
once             & 2 & 0 & $\checkmark$ \\
rwlock           & 7 & 0 & $\checkmark$ \\
wait             & 1 & 0 & -- \\
\midrule
\textbf{Total (19 modules)} & 42 & 33 & 17 \\
\bottomrule
\end{tabular}
\end{table}

\subsection{Task Extraction}

We selected source modules containing the target proof-obligation families and recursively inlined local dependencies into standalone Rust files. Focused tasks remove ghost code associated with selected atomic operations or invariant openings, or the body of a selected lemma, while preserving executable code and specifications. Each task targets one function, except \texttt{wait} and \texttt{cuckoo\_inductive}, which group functions with short, similar proof gaps.

This process yielded 75 focused tasks: 42 atomic-operation tasks and 33 state-machine tasks. For 17 source modules, we also removed all selected proofs simultaneously to create full-module variants requiring multiple repairs. These variants reuse the focused tasks' proof obligations, providing complementary evaluations of isolated and combined repairs rather than 92 independent obligations.

\section{Evaluation Framework}

The tasks are accompanied by an automated evaluation framework used to collect the results presented in Section~\ref{experimental_evaluation}.
Every agent must follow the same proof-repair contract. It may add proof code, but
it must preserve all executable code, specification code, and existing function
contracts. In particular, the agent may not weaken a precondition or
postcondition, replace executable statements, introduce an axiom, or mark an
existing function as trusted. The use of \texttt{admit} and \texttt{assume} is
also prohibited. This way, agents cannot obtain a
passing result by moving, removing, or replacing existing behavior.

\subsection{Execution}

Each task runs in a temporary directory inside a dedicated Docker container with the agent's CLI setup and a fixed base image built with the latest or another specified Verus version, its standard library (\texttt{vstd}), and Lynette~\cite{autoverus, verusage}, a cheat checker which we extended to support the corpus's proof forms. The prompt specifies input/output paths and edit restrictions, instructing the agent to verify its changes with Verus until a candidate passes verification and the cheat checker.

Since public repositories may expose reference proofs, external search and download tools are disabled or blocked, while model-provider access remains enabled. Session logs support manual inspection to identify and repeat runs that accessed external solutions.

Agents are given 20 minutes for each task. When this limit is reached, the
agent container is stopped and the task marked as timed out. If the agent
produces the expected output file, the evaluation framework extracts
and evaluates it; otherwise, the task records that no candidate solution was produced.

\subsection{Solution Validation and Recorded Outputs}

Each candidate is validated in a fresh container built from the benchmark
base image and accepted only if it passes Verus verification and Lynette.

The harness preserves generated solutions and complete session logs.
Agent-specific parsers extract estimated model cost, agent turns, tool
calls, and session duration. These statistics, validation results, timeout
status, and solution paths are recorded incrementally in a run-level JSON
file.

Infrastructure failures, including rate limits, abnormal exits, and invalid
logs, leave tasks pending for resumption without repeating completed work.

% Timeouts are final outcomes under the stated resource budget.

% Candidate validation is performed in a newly
% created container based only on the benchmark base image. 
% A generated file is accepted only if it passes Verus verification and the
% cheat checker, Lynette~\cite{autoverus, verusage}, which we extended to recognize the proof forms used by
% the corpus.

% For each run, the harness preserves the generated solution and the complete
% agent session log for later inspection. Agent-specific parsers process these logs into a
% common set of
% statistics: estimated model cost, number of agent turns, number of tool calls,
% and session duration. These values, together with the timeout flag,
% Verus result,
% change-checker result, and solution path, are stored in a run-level JSON file
% that is updated after every completed task. 

% Provider rate limits, abnormal agent exits, and invalid or incomplete session
% logs are treated as infrastructure failures rather than proof failures. Such
% tasks remain pending, and the incremental recordings allow resuming runs without repeating completed tasks.
% Timeouts are final task outcomes under the stated resource budget.

\section{Experimental Evaluation\label{experimental_evaluation}}

We evaluate four coding-agent harnesses on \benchmarkName{}\ to assess repair effectiveness, challenging proof patterns, and resource consumption: Codex~\cite{codex2026}, OpenCode~\cite{opencode2026}, GitHub Copilot~\cite{githubcopilot2026}, and Pi~\cite{pi2026}. All use medium reasoning effort and otherwise default configurations.

% We evaluated four coding-agent harnesses on \benchmarkName{} to assess repair effectiveness, identify challenging proof patterns, and examine the relationship between repair success and resource consumption. We compare Codex~\cite{codex2026}, OpenCode~\cite{opencode2026}, GitHub Copilot~\cite{githubcopilot2026}, and Pi~\cite{pi2026}. Each harness uses medium model reasoning effort and otherwise retains its default configuration.

\paragraph{\textbf{Proof-repair effectiveness}}

% \begin{table}[!htb]
% \caption{Initial valid repairs~\cite{ribeiro} in \benchmarkName{}}
% \label{tab:results}
% \centering
% \begin{tabular}{lrr}
% \toprule
% Harness & Focused (75) & Full-module (17) \\
% \midrule
% Codex & 62 (83\%) & 7 (41\%) \\
% GitHub Copilot & 56 (75\%) & 5 (29\%) \\
% OpenCode & 64 (85\%) & 8 (47\%) \\
% Pi & 63 (84\%) & 8 (47\%) \\
% \bottomrule
% \end{tabular}
% \end{table}

Agents repaired 56--64 of 75 focused tasks (75--85\%), versus 5--8 of 17 full-module tasks (29--47\%) with GPT-5.6 Luna~\cite{gpt56luna2026}. A follow-up using Claude Sonnet 4.6~\cite{sonnet462026} with OpenCode and Pi solved five of the seven focused tasks that all initial agents had failed, but only one of the seven remaining unsolved full-module tasks. These baselines suggest that coordinating multiple repairs within a module is harder than completing isolated proof gaps.

\paragraph{\textbf{Proof difficulty}}
All seven focused tasks initially unsolved by every agent involved atomic operations. Repair success declined with larger proof omissions and more atomic sites. Qualitative analysis revealed difficulties preserving arithmetic bounds, applying existing lemmas, and avoiding unproductive responses to verifier feedback. These findings illustrate the benchmark's potential for analysing proof difficulty and guiding concurrency-aware repair.

\paragraph{\textbf{Resource efficiency}}
The study compared repair success with token consumption, tool calls, and monetary cost. Pi achieved success rates among the highest observed while maintaining the lowest mean cost per attempt in both task sets. Across the evaluated configurations, greater resource consumption did not yield more successful repairs. The benchmark, therefore, supports comparisons of repair effectiveness and resource efficiency under a common evaluation protocol.

\begin{table*}[!tb]
  \centering
  \caption{Comparison of Verus benchmarks.}
  \label{tab:benchmark-comparison}
  \small
  \begin{tabularx}{\textwidth}{
    @{}l l
    >{\raggedright\arraybackslash}X
    >{\raggedright\arraybackslash}X@{}
  }
    \toprule
    \textbf{Benchmark} &
    \textbf{Size} &
    \textbf{Main task} &
    \textbf{Focus} \\
    \midrule
    Verus-Bench~\cite{autoverus} &
    150 tasks &
    Proof repair &
    Small algorithm programs \\

    VeruSAGE-Bench~\cite{verusage} &
    849 tasks &
    Proof repair &
    Real systems code \\

    VeriStruct~\cite{veristruct} &
    11 modules &
    Specification + proof synthesis &
    Data structures \\

    AlgoVeri~\cite{algoveri} &
    77 problems &
    Code + proof generation &
    Textbook algorithms \\

    VeriContest~\cite{vericontest} &
    946 problems &
    Specification, code, proof generation &
    Competitive programming \\

    LeetCode-Verus~\cite{leetcode-verus-repo} &
    37 examples &
    Verified program collections &
    LeetCode algorithms \\

    HumanEval-Verus~\cite{humanevalverus} &
    85 function tasks &
    Verified program collections &
    Small programming problems \\
    \midrule
    \textbf{VerusWeave} &
    \textbf{75 tasks / 17 modules} &
    \textbf{Concurrent proof repair} &
    \textbf{Atomics and tokenized state machines} \\
    \bottomrule
  \end{tabularx}
\end{table*}

\section{Related Work}

Table~\ref{tab:benchmark-comparison} compares existing Verus benchmarks by size, synthesis objectives, and coverage of program domains and Verus primitives. Despite their diversity, concurrency-specific proof obligations remain underrepresented. \benchmarkName{} addresses this gap by evaluating agents' ability to repair proofs involving two largely unexplored Verus concurrency primitives: permissioned atomics and tokenized state machines.

\section{Threats to Validity and Future Work}

\benchmarkName{} samples four projects rather than representative Rust concurrency code, reflecting the scarcity of public projects using these interfaces. As tasks originate from public repositories and recent models are evaluated, prior exposure to source proofs cannot be excluded.

Future extensions include larger proof holes and logical atomicity and weak-memory reasoning. Agent involvement could also expand beyond proof repair to executable-code generation alongside proofs, guided by behavioural specifications.
% Porting and verifying
% new concurrent data structures in Verus would provide a strong test of whether
% coding agents can contribute throughout the verification process
% rather than only
% complete proof fragments within an established design.

\section{Conclusion}

We presented \benchmarkName{}, the first Verus benchmark focused entirely on concurrent code, with 75 focused proof-repair tasks and 17 full-module variants from four open-source projects. It covers ghost-resource reasoning around atomic operations and invariant preservation in tokenized state machines. Its evaluation framework validates repairs and records agent behaviour and resource use. Initial results show higher success on focused tasks than full-module variants, motivating further evaluation of extended concurrency proof repairs. \benchmarkName{} provides a reusable basis for evaluating and improving automated proof construction and verification tools for concurrent software.

\section*{Acknowledgements}

This work was co-funded by the European Union through the Lisboa 2030 Programme (ERDF) and by national funds through FCT, I.P., under project no. 15018 (\href{https://safeiac.github.io}{project/ SafeIaC}, DOI: \href{https://doi.org/10.54499/2023.18089.ICDT}{10.54499/2023.18089.ICDT}).

\bibliographystyle{IEEEtran}
\bibliography{references}

@misc{brugger2025practicaldeductiveverificationinsights,
      title={Toward Practical Deductive Verification: Insights from a Qualitative Survey in Industry and Academia},
      author={Lea Salome Brugger and Xavier Denis and Peter Müller},
      year={2025},
      eprint={2510.20514},
      archivePrefix={arXiv},
      primaryClass={cs.SE},
      url={https://arxiv.org/abs/2510.20514},
}

@article{lattuada2023verusverifyingrustprograms,
      title = {{Verus}: Verifying {Rust} Programs using Linear Ghost Types},
      author = {Andrea Lattuada and Travis Hance and Chanhee Cho and Matthias Brun and Isitha Subasinghe and Yi Zhou and Jon Howell and Bryan Parno and Chris Hawblitzel},
      journal = {Proceedings of the ACM on Programming Languages},
      volume = {7},
      number = {OOPSLA1},
      year = {2023},
      pages = {286--315},
      articleno = {85},
      numpages = {30},
      publisher = {Association for Computing Machinery},
      doi = {10.1145/3586037},
      url = {https://doi.org/10.1145/3586037},
}

@inproceedings{cortenmm,
    author = {Zhang, Junyang and Xu, Xiangcan and Zou, Yonghao and Tang, Zhe and Wan, Xinyi and Hu, Kang and Wang, Siyuan and Xu, Wenbo and Wang, Di and Chen, Hao and Huang, Lin and Yan, Shoumeng and Tamir, Yuval and Luo, Yingwei and Wang, Xiaolin and Yu, Huashan and Wang, Zhenlin and Tian, Hongliang and Zhou, Diyu},
    title = {{CortenMM}: Efficient Memory Management with Strong Correctness Guarantees},
    year = {2025},
    isbn = {9798400718700},
    publisher = {Association for Computing Machinery},
    address = {New York, NY, USA},
    url = {https://doi.org/10.1145/3731569.3764836},
    doi = {10.1145/3731569.3764836},
    booktitle = {Proceedings of the ACM SIGOPS 31st Symposium on Operating Systems Principles},
    pages = {1082–1098},
    numpages = {17},
    location = {Lotte Hotel World, Seoul, Republic of Korea},
    series = {SOSP '25}
}

@inproceedings {verismo,
    author = {Ziqiao Zhou and Anjali and Weiteng Chen and Sishuai Gong and Chris Hawblitzel and Weidong Cui},
    title = {{VeriSMo}: A Verified Security Module for Confidential {VMs}},
    booktitle = {18th USENIX Symposium on Operating Systems Design and Implementation (OSDI 24)},
    year = {2024},
    isbn = {978-1-939133-40-3},
    address = {Santa Clara, CA},
    pages = {599--614},
    url = {https://www.usenix.org/conference/osdi24/presentation/zhou},
    publisher = {USENIX Association},
    month = jul
}

@misc{owlc,
      author = {Pratap Singh and Joshua Gancher and Bryan Parno},
      title = {{OwlC}: Compiling Security Protocols to Verified, Secure, High-Performance Libraries},
      howpublished = {Cryptology {ePrint} Archive, Paper 2025/1092},
      year = {2025},
      url = {https://eprint.iacr.org/2025/1092}
}

@article{autoverus,
   title={{AutoVerus}: Automated Proof Generation for {Rust} Code},
   volume={9},
   ISSN={2475-1421},
   url={http://dx.doi.org/10.1145/3763174},
   DOI={10.1145/3763174},
   number={OOPSLA2},
   journal={Proceedings of the ACM on Programming Languages},
   publisher={Association for Computing Machinery (ACM)},
   author={Yang, Chenyuan and Li, Xuheng and Misu, Md Rakib Hossain and Yao, Jianan and Cui, Weidong and Gong, Yeyun and Hawblitzel, Chris and Lahiri, Shuvendu and Lorch, Jacob R. and Lu, Shuai and Yang, Fan and Zhou, Ziqiao and Lu, Shan},
   year={2025},
   month=oct, pages={3454–3482}
}

@inproceedings{veristruct,
  title={{VeriStruct}: {AI}-assisted Automated Verification of Data-Structure Modules in {Verus}},
  author={Sun, Chuyue and Sun, Yican and Amrollahi, Daneshvar and Zhang, Ethan and Lahiri, Shuvendu and Lu, Shan and Dill, David and Barrett, Clark},
  booktitle={Proceedings of the 32nd International Conference on Tools and Algorithms for the Construction and Analysis of Systems (TACAS)},
  year={2026}
}

@misc{verusage,
  title={{VeruSAGE}: A Study of Agent-Based Verification for {Rust} Systems},
  author={Chenyuan Yang and Natalie Neamtu and Chris Hawblitzel and Jacob R. Lorch and Shan Lu},
  year={2025},
  eprint={2512.18436},
  archivePrefix={arXiv},
  primaryClass={cs.OS},
  url={https://arxiv.org/abs/2512.18436},
}

@misc{humanevalverus,
  author       = {Alex Bai and Jay Bosamiya and Edwin Fernando and Md Rakib Hossain and Jay Lorch and Shan Lu and Natalie Neamtu and Bryan Parno and Amar Shah and Elanor Tang},
  title        = {{HumanEval-Verus}: Hand-Written Examples of Verified {Verus} Code Derived From {HumanEval}},
  year         = {2024},
  howpublished = {\url{https://github.com/secure-foundations/human-eval-verus}},
  note         = {GitHub repository}
}

@misc{verified_memory_allocator,
  author       = {{Verus Project Contributors}},
  title        = {verified-memory-allocator, based on mimalloc},
  year         = {2024},
  howpublished = {\url{https://github.com/verus-lang/verified-memory-allocator}},
  note         = {Accessed: 2026-09-07},
}

@software{verified_node_replication,
  author       = {{Verus Project Contributors}},
  title        = {verified-node-replication},
  year         = {2024},
  howpublished = {\url{https://github.com/verus-lang/verified-node-replication}},
  note         = {Accessed: 2026-09-07},
}

@software{vostd_repo,
  author       = {{VOSTD Project Contributors}},
  title        = {vostd},
  year         = {2025},
  howpublished = {\url{https://github.com/asterinas/vostd}},
  note         = {Accessed: 2026-09-07},
}

@software{verus_repo,
  author       = {{Verus Project Contributors}},
  title        = {verus},
  year         = {2024},
  howpublished = {\url{https://github.com/verus-lang/verus}},
  note         = {Accessed: 2026-09-07},
}

@software{asterinas,
  author  = {{Asterinas Project}},
  title   = {{Asterinas}},
  url     = {https://github.com/asterinas/asterinas},
  version = {commit SHA or release version},
  urldate = {2026-09-07}
}

@inproceedings{oliveira2025challenges,
  title={What Challenges Do Developers Face When Using Verification-Aware Programming Languages?},
  author={Oliveira, Francisco and Mendes, Alexandra and Carreira, Carolina},
  booktitle={36th IEEE International Symposium on Software Reliability Engineering (ISSRE)},
  year={2025}
}

@article{leveraging_llms_dafny,
  title={Leveraging Large Language Models to Boost {Dafny}’s Developers Productivity},
  author={{\'A}lvaro F. Silva and Alexandra Mendes and Jo{\~a}o F. Ferreira},
  journal={2024 IEEE/ACM 12th International Conference on Formal Methods in Software Engineering (FormaliSE)},
  year={2024},
  pages={138-142},
  url={https://api.semanticscholar.org/CorpusID:266725333}
}

@phdthesis{hance2024verifying,
    author = {Travis Hance},
    title = {Verifying Concurrent Systems Code},
    school = {Carnegie Mellon University},
    year = {2024},
    month = aug,
    number = {CMU-CS-24-146}
}

@misc{opencode2026,
    author       = {{OpenCode}},
    title        = {{OpenCode}: The Open Source {AI} Coding Agent},
    year         = {2026},
    howpublished = {\url{https://opencode.ai/en/docs}},
    note         = {Accessed: 2026-09-02}
}

@misc{codex2026,
    author       = {{OpenAI}},
    title        = {{OpenAI} {Codex} {CLI}},
    year         = {2026},
    howpublished = {\url{https://github.com/openai/codex}},
    note         = {Accessed: 2026-09-02}
}

@misc{gpt56luna2026,
    author       = {{OpenAI}},
    title        = {{GPT-5.6 Luna Model}},
    year         = {2026},
    howpublished = {\url{https://developers.openai.com/api/docs/models/gpt-5.6-luna}},
    note         = {Accessed: 2026-09-08}
}

@misc{sonnet462026,
    author       = {{Anthropic}},
    title        = {Introducing {Claude Sonnet} 4.6},
    year         = {2026},
    month        = feb,
    howpublished = {\url{https://www.anthropic.com/news/claude-sonnet-4-6}},
    note         = {Accessed: 2026-09-08}
}

@misc{githubcopilot2026,
    author       = {{GitHub}},
    title        = {{GitHub Copilot}},
    year         = {2026},
    howpublished = {\url{https://docs.github.com/en/copilot/get-started/what-is-github-copilot}},
    note         = {Accessed: 2026-09-02}
}

@misc{pi2026,
    author       = {{Earendil Inc. and Contributors}},
    title        = {{Pi}: A Minimal Agent Harness},
    year         = {2026},
    howpublished = {\url{https://pi.dev/}},
    note         = {Accessed: 2026-09-02}
}

@InProceedings{algoveri,
  title = 	 {{AlgoVeri}: An Aligned Benchmark for Verified Code Generation on Classical Algorithms},
  author =       {Zhao, Haoyu and Yang, Ziran and Li, Jiawei and He, Mike and Li, Zenan and Jin, Chi and Veeravalli, Venugopal and Gupta, Aarti and Arora, Sanjeev},
  booktitle = 	 {Proceedings of the 43rd International Conference on Machine Learning},
  pages = 	 {162621--162652},
  year = 	 {2026},
  editor = 	 {Zhang, Tong and Dudik, Miroslav and Jaggi, Martin and Agarwal, Alekh and Li, Sharon and Schuurmans, Dale and Zhu, Jerry and Berkenkamp, Felix and Dong, Hanze and Bietti, Alberto},
  volume = 	 {306},
  series = 	 {Proceedings of Machine Learning Research},
  month = 	 {06--11 Jul},
  publisher =    {PMLR},
  url = 	 {https://proceedings.mlr.press/v306/zhao26bm.html}
}

@misc{vericontest,
    title={{VeriContest}: A Competitive-Programming Benchmark for Verifiable Code Generation},
    author={Zichen Xie and Mrigank Pawagi and Yuxin Liu and Aaditi Rai and Lize Shao and John Berberian Jr. and Sicong Che and Wenxi Wang},
    year={2026},
    eprint={2605.08553},
    archivePrefix={arXiv},
    primaryClass={cs.SE},
    url={https://arxiv.org/abs/2605.08553},
}

@misc{leetcode-verus-repo,
    author       = {Weituo Dai and Contributors},
    title        = {{Verus} Study Cases: {LeetCode}},
    year         = {2025},
    howpublished = {\url{https://github.com/WeituoDAI/verus-study-cases-leetcode}},
    note         = {GitHub repository, accessed: 2026-09-20}
}

@article{rango,
  title={{Rango}: Adaptive Retrieval-Augmented Proving for Automated Software Verification},
  author={Kyle Thompson and Nuno Saavedra and Pedro Carrott and Kevin Fisher and Alex Sanchez-Stern and Yuriy Brun and Jo{\~a}o F. Ferreira and Sorin Lerner and Emily First},
  journal={2025 IEEE/ACM 47th International Conference on Software Engineering (ICSE)},
  year={2024},
  pages={347-359},
  url={https://api.semanticscholar.org/CorpusID:274710349}
}

@article{autorocq,
author = {Tu, Haoxin and Zhao, Huan and Song, Yahui and Zafar, Mehtab and Meng, Ruijie and Roychoudhury, Abhik},
title = {Agentic Verification of Software Systems},
year = {2026},
issue_date = {July 2026},
publisher = {Association for Computing Machinery},
address = {New York, NY, USA},
volume = {3},
number = {FSE},
url = {https://doi.org/10.1145/3808164},
doi = {10.1145/3808164},
journal = {Proc. ACM Softw. Eng.},
month = jun,
articleno = {FSE157},
numpages = {24}
}

@inproceedings{chess,
author = {Musuvathi, Madanlal and Qadeer, Shaz and Ball, Thomas and Basler, Gerard and Nainar, Piramanayagam Arumuga and Neamtiu, Iulian},
title = {Finding and reproducing {Heisenbugs} in concurrent programs},
year = {2008},
publisher = {USENIX Association},
address = {USA},
booktitle = {Proceedings of the 8th USENIX Conference on Operating Systems Design and Implementation},
pages = {267–280},
numpages = {14},
location = {San Diego, California},
series = {OSDI'08}
}

@article{silva2025inferring,
  title={Inferring multiple helper Dafny assertions with LLMs},
  author={Silva, {\'A}lvaro and Mendes, Alexandra and Martins, Ruben},
  journal={arXiv preprint arXiv:2511.00125},
  year={2025}
}

\end{document}